\documentclass[twocolumn,10pt]{article}

\usepackage{epsf,a4,cite,xspace}

\title{Diffusion of colloids at short times
       }
\author{M. Watzlawek\thanks{
              Present address: 
              Institut f\"ur Theoretische Physik II, 
              Heinrich--Heine--Universit\"at, Universit\"atsstr. 1, 
              D--40225 D\"usseldorf, Germany. 
              Email: martin@thphy.uni-duesseldorf.de
           } and G. N\"agele \\ \hfill \\
           {\small Fakult\"at f\"ur Physik, Universit\"at Konstanz,} \\
           {\small Postfach 5560,
           D--78434 Konstanz, Germany}}
\date{September 26, 1996 \\
        {\small [published in Prog. Coll. Polym. Sci. {\bf 104}, 168 (1997)]}
     }
\begin{document}

\maketitle

\newcommand{\ve}[1]{\mbox{\boldmath$#1$}}
\newcommand{\tr}{\mbox{Tr}}
\newcommand{\av}[1]{\mbox{$\left\langle #1\right\rangle$}}


\begin{abstract}
   We study the combined effects of electrostatic and 
   hydrodynamic interactions (HI) on the 
   short-time dynamics of charge-stabilized colloidal spheres.
   For this purpose, we calculate 
   the translational and the rotational self-diffusion
   coefficients, $D^t_s$ and $D^r_s$, as function of volume fraction
   $\phi$ for various values of the 
   effective particle charge $Z$ and various concentrations
   $n_s$ of added 1--1 electrolyte. 

   Our results show that the self-diffusion coefficients
   in deionized suspensions are less affected by HI than in 
   suspensions with added electrolyte. For very large $n_s$,
   we recover the well-known results for
   hard spheres, i.e. a linear $\phi$-dependence of
   $D^t_s$ and $D^r_s$ at small $\phi$. In contrast, 
   for deionized charged suspensions at small $\phi$,
   we observe the interesting non-linear scaling properties 
   $D^t_s\propto1-a_t\phi^{4/3}$ and $D^r_s\propto 1-a_r\phi^2$.
   The coefficients $a_t$ and $a_r$ are found to be nearly 
   independent of $Z$. 
   The qualitative differences between the dynamics of charged and uncharged
   particles can be well explained in terms of an
   effective hard sphere (EHS) model.
\end{abstract}

\section{Introduction}

Since several years, the effect of HI on the 
short-time self-diffusion
coefficients of hard sphere suspensions has been
investigated in detail by various authors
\cite{Cichocki:88,Jones:1:88,Jones:Pusey:91,Beenakker:Mazur:83,Degiorgio:95}.
For the calculation of the first and second virial coefficients of
$D^t_s$ and $D^r_s$ in an expansion in terms of the volume fraction $\phi$, 
both 
the influence of two-body and three-body HI
was taken into account.
At small $\phi$, the
currently established results for the normalized diffusion coefficients 
$H^t_s$ and $H^r_s$ are given by \cite{Cichocki:88,Beenakker:Mazur:83}
\begin{equation} \label{hts.hs}
   H^t_s=\frac{D^t_s}{D^t_0}=1-1.831\phi+0.88\phi^2+\mathcal{O}(\phi^3)
\end{equation}
and by \cite{Jones:1:88,Degiorgio:95}   
\begin{equation} \label{hrs.hs}
   H^r_s=\frac{D^r_s}{D^r_0}=1-0.630\phi-0.67\phi^2+\mathcal{O}(\phi^3),
\end{equation}
respectively.
Here, $D^t_0$ and $D^r_0$ are the  
Stokesian diffusion coefficients for a colloidal sphere of radius $a$
dispersed
in a solvent of viscosity $\eta$.

The possibility to express $H^t_s$ and $H^r_s$ in terms of a 
power series in $\phi$ arises from the fact that  hard sphere suspensions
at small $\phi$ can be considered as dilute both with respect
to the particle hydrodynamics and
to the microstructure.
For charge-stabilized suspensions, however, this is not possible in general 
\cite{Naegele:Habil:published,Watzlawek:96:1}. 
Especially deionized, i.e. salt-free suspensions
exhibite pronounced spatial correlations even at very small $\phi$, 
so that these systems are diluted only as far as the HI is concerned.
The corresponding
radial distribution function $g(r)$ has 
a pronounced
$\phi$-dependence, a well developed first maximum, 
and it shows a correlation hole, i.e. a sperical region with zero
probability for finding another particle, which usually
extends over several
particle diameters \cite{Naegele:Habil:published,Watzlawek:96:1}. 
In contrast, the $g(r)$ of hard spheres is nearly
a unit step function $g(r)\simeq\Theta(r-2a)$ for $\phi\le0.05$. 
Therefore, the calculation
of $H^t_s$ and $H^r_s$ at small $\phi$ is more demanding
for charged suspensions 
than for hard spheres, because for the charged particles 
it is necessary to use
distributions functions generated from
computer simulations or 
integral equation methods 
\cite{Naegele:Habil:published,Watzlawek:96:1}.

We will show subsequently, that it is essentially the presence 
of the correlation hole for charged suspensions, which
causes large and interesting differences in
the short-time diffusion of
charged and uncharged suspensions.

\section{Calculation of {\boldmath $H^t_s$} and {\boldmath $H^r_s$}}

In the following, we shortly summarize the main expressions 
needed to calculate
$H^t_s$ and $H^r_s$ for charge-stabilized suspensions.
A more detailed description of the method used by us for the calculation
of short-time diffusion coefficients 
is given in Refs. \cite{Watzlawek:96:1,Watzlawek:forthcoming}.

As shown in Refs. \cite{Degiorgio:94,Degiorgio:95}, both
$H^t_s$ and $H^r_s$ can be  measured 
using depolarized dynamic light scattering (DDLS)
from suspensions of optically anisotropic colloidal spheres.
On the time scales, which are accessable by DDLS, the theoretical expression
for
$H^t_s$ is given by \cite{Jones:Pusey:91}
\begin{equation} \label{hts}
   H^t_s=\frac{1}{3D^t_0}\av{\tr \ve{D}^{tt}_{11}(\ve{r}^N)}.
\end{equation}
The corresponding expression for $H^r_s$ 
is obtained from eq. (\ref{hts}) by simply replacing the
superscript '$t$' by '$r$'.
The hydrodynamic diffusivity tensors $\ve{D}^{tt}_{11}$  and 
$\ve{D}^{rr}_{11}$
relate the force/torque exerted by the solvent
on an arbitrary particle $1$ with
its translational/angular velocity \cite{Jones:Pusey:91,Jones:Schmitz:88}.
Due to the many-body character of HI, both tensors depend on the
particle configuration
$\ve{r}^N=(\ve{r}_1,\ldots\ve{r}_N)$ of all $N$ interacting particles, 
and in principle the full $N$-particle 
distribution function is needed to perform the ensemble average 
$\av{\ldots}$. $\tr\ve{D}^{tt}_{11}$ denotes the sum 
over the diagonal elements of
$\ve{D}^{tt}_{11}$.

For an appropriate evaluation of eq. (\ref{hts}), we use
a rooted cluster expansion \cite{Jones:1:88,Degiorgio:95}, which leads
to a "hydrodynamic virial expansion" of $H^t_s$: 
\begin{equation} 
   \label{hydrodynamic.virial.expansion}
   H^t_s=1+H^t_{s1}\phi+H^t_{s2}\phi^2+\mathcal{O}(\phi^3).
\end{equation}
Here, the coefficient $H^t_{s1}$ is given by an
integral over the product of $g(r)$ with a translational hydrodynamic 
mobility function, which depends only on the distance $r$ of two spheres
\cite{Jones:1:88,Cichocki:88,Watzlawek:96:1,Watzlawek:forthcoming}.
The second coefficient $H^t_{s2}$ accounts for three-body HI. 
For evaluating $H^t_{s2}$, one needs therefore
an expression for the 
static triplett correlation function $g^{(3)}(\ve{r},\ve{r}')$ 
which appears as part of the integrand of a three-fold integral. 

A similar analysis is used to calculate $H^r_s$, leading to results
which involve now rotational hydrodynamic two-body and three-body 
mobility functions 
\cite{Degiorgio:95,Beenakker:Mazur:83,Watzlawek:96:1,Watzlawek:forthcoming}.

The results for $H^t_s$ and $H^r_s$ depicted
in eq. (\ref{hts.hs}) and (\ref{hrs.hs})
were derived from eq. (\ref{hts})
by using in $H^t_{s1}$ and $H^r_{s1}$
the $g(r)$ of hard spheres evaluated up to linear
order in $\phi$, whereas the vanishing density form of 
$g^{(3)}(\ve{r},\ve{r}')$ was used in calculating the coefficients
$H^t_{s2}$ and $H^r_{s2}$. 
In these results, 
exact two-body HI is accounted for $H^t_{s1}$ and $H^r_{s1}$, 
whereas only the leading long-distance contribution to the three-body
mobility functions  
was used for the calculation of $H^t_{s2}$ and $H^r_{s2}$
\cite{Cichocki:88,Jones:1:88,Beenakker:Mazur:83,Degiorgio:95}.

For charged suspensions, however, it is not possible to use in eq. (\ref{hts})
low-order virial expressions of the two-body and three-body
static 
distribution functions. In this study,
we use instead results for $g(r)$, which are
obtained from the rescaled mean spherical approximation
(RMSA), as applied to the one-component macrofluid model
of charge-stabilized colloidal suspensions \cite{Naegele:Habil:published}. 
The effective pair potential $u(r)$ acting between two particles
is modelled by the repulsive part of
the famous DLVO-potential, i.e. 
$\beta u(r)=K\exp\left[-\kappa(r-2a)\right]\frac{a}{r}$,
for $r>2a$. Here, $\beta=(k_BT)^{-1}$, 
$K=Z^2(L_B/a)(1+\kappa a)^{-2}$, $L_B=\beta e^2/\epsilon$, and
$\epsilon$ denotes the dielectric constant of the solvent.
The screening parameter $\kappa$ is given by
$\kappa^2=L_B\left[3|Z|\phi/a^3+8\pi n_s\right]$, where $n_s$ is 
the concentration of added 1--1 electrolyte, and the counterions are assumed
to be monovalent
\cite{Naegele:Habil:published,Watzlawek:96:1}.
Moreover, we use Kirkwood's
superposition approximation for 
$g^{(3)}(\ve{r},\ve{r}')$, inserting again the RMSA--$g(r)$.
Further details concerning the numerical calculation
of
$H^t_{s1}$, $H^r_{s1}$, $H^t_{s2}$, and $H^r_{s2}$ are given in Refs. 
\cite{Watzlawek:96:1,Watzlawek:forthcoming}. 

\section{Results and discussion}

\begin{figure}
     \epsfxsize=7.5cm
     \epsfysize=4.8cm
     \hfill\epsfbox{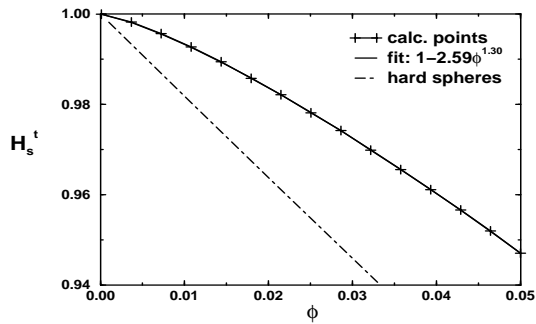}\hfill~
      \caption{ \label{trans.nosalt.fit.large.plot} \small
               Normalized short-time translational diffusion
               coefficient $H^t_s$ for a deionized charged suspension with
               $Z=200$, $a=45$nm, $T=294$K, and $\epsilon=87.0$.
               Also shown is the result for hard spheres according to
               eq. (\ref{hts.hs}).
              }
\end{figure}

\begin{figure}
      \epsfxsize=7.5cm
      \epsfysize=4.8cm
      \hfill\epsfbox{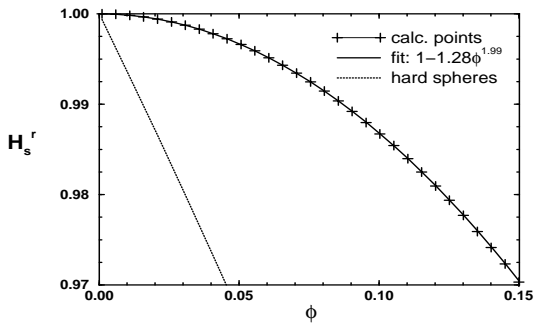}\hfill~
      \caption{ \label{rot.nosalt.fit.plot} \small
               Normalized short-time rotational diffusion coefficient
               $H^r_s$ for a deionized suspension with
               system parameters as in fig. \ref{trans.nosalt.fit.large.plot}.
               Also displayed  is the result for hard spheres according to
               eq. (\ref{hrs.hs}).
              }
\end{figure}

We focus first on the short-time diffusion coefficients of
deionized charged suspensions, i.e. where $n_s=0$.
Our results for $H^t_s$ and $H^r_s$ are
shown in figs. 
\ref{trans.nosalt.fit.large.plot} and \ref{rot.nosalt.fit.plot}.
The used system parameters are typical for systems which have been
under experimental
study \cite{Bitzer:private}.
Obviously, the effect of HI on the self-diffusion coefficients 
is less pronounced for charged suspensions than for hard spheres at the 
same $\phi$.
Furthermore, we find a quite different volume fraction dependence 
of $H^t_s$ and $H^r_s$ for charged and uncharged particles.
Whereas for hard spheres 
the $\phi$-dependence of $H^t_s$ and $H^r_s$ is linear
at small $\phi$ (cf. eqs. (\ref{hts.hs}) and (\ref{hrs.hs})), 
we obtain from a least-square fit of our numerical results 
(shown as crosses in figs. \ref{trans.nosalt.fit.large.plot}
and \ref{rot.nosalt.fit.plot})
the following results 
for deionized charged suspensions for $0\le\phi\le0.05$
\cite{Watzlawek:forthcoming}
\begin{eqnarray}
   \label{hts.charged}
   H^t_s&=&1-a_t\phi^{1.30}, \ \ a_t=2.59,
\\
   \label{hrs.charged}
   H^r_s&=&1-a_r\phi^{1.99}, \ \ a_r=1.28.
\end{eqnarray}
The coefficients $a_t$ and $a_r$
are found to be nearly independent of the effective particle
charge when $Z\ge200$ \cite{Watzlawek:96:1,Watzlawek:forthcoming}.
Note from fig. \ref{rot.nosalt.fit.plot}, that eq. (\ref{hrs.charged})
constitutes the best fit function for $H^r_s(\phi)$ 
even in the extended interval $0\le\phi\le0.15$.
In case of $H^t_s$ however, the parametric form
$H^t_s=a_t\phi^p$ provides no good fit for values of $\phi$ extending
beyond $0.05$
\cite{Watzlawek:forthcoming}.

There is a simple physical explanation for the weaker influence of HI on the
self-diffusion 
coefficients of charged suspensions as compared to uncharged ones.
As already mentioned, the $g(r)$ of 
deionized suspensions displayes a pronounced
correlation hole, resulting form
the strong electrostatic interparticle repulsion. 
Consequently,  
the hydrodynamic coupling between
the translational or rotational motions of two spheres becomes
rather small, thus giving rise
to the observed weak influence of HI. 
Unlike charged particles,
the influence particularly of the short-range part of
HI is rather strong for hard sphere suspensions at small $\phi$. This is due
to the large probability of finding hard sphere particles at contact 
or close to the contact distance $r=2a$.

Along this type of 
arguments, it is also possible to explain the differences of
$H^t_s(\phi)$ and $H^r_s(\phi)$ in deionized suspensions and suspensions with
nonvanishing 
$n_s$. We only show here the 
results of our calculations of $H^r_s(\phi)$ for example. 
\begin{figure}
      \epsfxsize=7.5cm
      \epsfysize=4.8cm
      \hfill\epsfbox{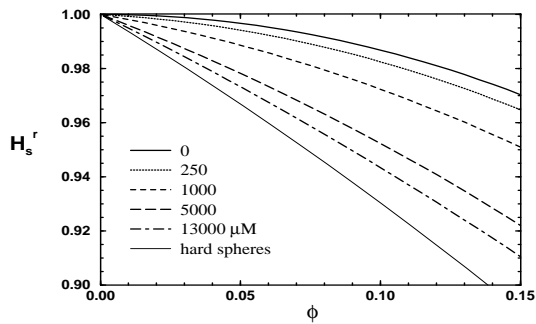}\hfill~
      \caption{ \label{rot.withsalt.plot} \small
               Volume fraction dependence of
               $H^r_s$ for various amounts of added 1--1
               electrolyte, as indicated in the figure. All other
               system parameters as in fig.\ref{trans.nosalt.fit.large.plot}.
              }
\end{figure}  
From these results in fig. \ref{rot.withsalt.plot}, we notice 
that $H^r_s$ becomes more 
and more
affected by HI when $n_s$ is increased.
For very large $n_s$, $H^r_s$ of charged particles approches
the result for hard sphere suspensions,
obtained semianalytically in Ref. \cite{Degiorgio:95} 
(cf. eq. (\ref{hrs.hs})).

This finding is easily explained
by noticing, that the extension of the 
correlation hole decreases with increasing $n_s$, leading
to a stronger hydrodynamic coupling of the particles.
Upon addition of electrolyte, the electrostatic repulsion
of the particles becomes more and more screened and 
short-ranged, resulting
in a pure hard-core repulsion for $n_s\to\infty$
\cite{Naegele:Habil:published,Watzlawek:96:1,Watzlawek:forthcoming}. 
Therefore, the microstructure of the suspension gradually transforms
to that of hard spheres, with $H^r_s$ 
approching the parametric form given in eq. (\ref{hrs.hs}).
We further note, that the radial distribution function
$g(r)$ corresponding to fig. \ref{rot.withsalt.plot} exhibites
a small correlation hole even for $n_s=13mM$.
This leads to the small differences of our results for $n_s=13mM$ and 
for hard spheres in fig. \ref{rot.withsalt.plot}.
 
We mention, that 
our results for $H^t_s$ show similar trends, i.e.
a gradual transformation of the $\phi$-dependence of $H^t_s$ from
eq. (\ref{hts.charged}) to eq. (\ref{hts.hs}) with
increasing $n_s$ \cite{Watzlawek:forthcoming}.

In the remainder of this article, we focus on the 
qualitatively different
$\phi$-dependencies of $H^t_s$ and $H^r_s$ found in case of
deionized charged
and uncharged suspensions.
For an intuitive physical explanation, we use an effective hard sphere
model (EHS model) \cite{Watzlawek:96:1,Naegele:Habil:published}, 
describing the actual $g(r)$ as
a unit step function $g_{EHS}(r)=\Theta(r-2a_{EHS})$. The EHS 
radius $a_{EHS}>a$ accounts in a crude fashion for the
correlation hole, observed in the actual $g(r)$. 
We identify $2a_{EHS}=r_m$, where $r_m$ is the position
of the first maximum of $g(r)$.
It is now crucial to notice that $r_m$ 
shows an interesting scaling property when $n_s=0$. 
Due to the strong electrostatic repulsion, $r_m$ has
the same $\phi$-dependence as the average geometrical distance
$\bar{r}$ between two spheres. Hence
\begin{equation} \label{scaling}
   a_{EHS}\propto r_{m}\propto
      \bar{r}=a\sqrt[3]{4\pi/3}\phi^{-\frac{1}{3}}.
\end{equation}
Using the approximation $g_{EHS}(r)$ of $g(r)$, it is 
easy to calculate the coefficients
$H^t_{s1}$ and $H^r_{s1}$ in an approximative way.
By using far-field expansions of the hydrodynamic
two-body mobility functions 
\cite{Jones:Schmitz:88,Watzlawek:96:1,Watzlawek:forthcoming}, 
one obtains the following results from
the leading terms of these expansions
\begin{eqnarray}
   H^t_{s1}&=&-\frac{15}{8}\left(\frac{a}{a_{EHS}}\right)
              +\mathcal{O}(a_{EHS}^{-3}),
\\
   H^r_{s1}&=&-\frac{5}{16}\left(\frac{a}{a_{EHS}}\right)^3
              +\mathcal{O}(a_{EHS}^{-5}).
\end{eqnarray}  
This leads together with eq. (\ref{hydrodynamic.virial.expansion}) and
(\ref{scaling}) to the expressions 
\begin{eqnarray}
   \label{hts.ehs}
   H^t_s&=&1-A^t\phi^{\frac{4}{3}}+\mathcal{O}(\phi^2), \ \ A^t>0,
\\
   \label{hrs.ehs}
   H^r_s&=&1-A^r\phi^2+\mathcal{O}(\phi^{\frac{8}{3}}), \ \ A^r>0,
\end{eqnarray}  
with exponents which are in good agreement with our numerical findings given
in figs. \ref{trans.nosalt.fit.large.plot} and \ref{rot.nosalt.fit.plot}
(cf. eqs. (\ref{hts.charged}) and (\ref{hrs.charged})).

Therefore we have shown by a simple analytic calculation based on
the EHS model, 
that the observed differences in
the functional forms of $H^t_s(\phi)$ and $H^r_s(\phi)$ between charged
and uncharged suspensions are mainly caused by the leading terms
of the hydrodynamic two-body mobility functions in combination with the
scaling property $r_m\propto\phi^{-1/3}$, valid for
deionized suspensions. The higher order terms in the
hydrodynamic far-field expansions only give rise to minor corrections
to the observed scaling properties depicted
in eq. (\ref{hts.charged}) and
(\ref{hrs.charged}). These terms become increasingly
important for larger volume fractions 
$\phi\ge0.05$ (cf. eqs. (\ref{hts.ehs}) and (\ref{hrs.ehs}) in the
EHS model). 

When electrolyte is added to the suspension, eq. (\ref{scaling}) becomes
invalid because of the enhanced screening of
the direct particle interactions. This  
causes a change in the functional behaviour of
$H^t_s(\phi)$ and $H^r_{s}(\phi)$, as can be seen both from
the EHS model and from our numerical results 
(cf. fig. \ref{rot.withsalt.plot} in case of $H^r_s$) 
\cite{Watzlawek:96:1,Watzlawek:forthcoming}.
  
Using the EHS model, it is also possible the motivate
the nearly
$Z$-independence of $a_t$ and $a_r$ in eqs. (\ref{hts.charged}) and
(\ref{hrs.charged}). Since 
$r_{m}$ is nearly independent of $Z$ for $Z\ge200$, the EHS model 
predicts charge independent results for the short-time diffusion
coefficients of deionized suspensions, in agreement with our numerical
results.

We mention, that it is also possible to deal with
$H^t_{s2}$ and
$H^r_{s2}$ within the EHS model, giving further insight in
the volume fraction dependence of the diffusion coefficients
of deionized suspensions \cite{Watzlawek:96:1,Watzlawek:forthcoming}.
It is then possible to explain qualitatively 
the surprising fact that
$H^r_s(\phi)$ is well parametrized up to $\phi=0.15$
by the functional form
$H^r_s=1-a_r\phi^2$, obtained in the EHS model by using only the
leading term in the far-field expansion of the rotational 
two-body  mobility functions \cite{Watzlawek:96:1,Watzlawek:forthcoming}.
 
\section{Conclusion}

We have presented calculations of the translational and
rotational short-time self-diffusion coefficients for
charge-stabilized suspensions. The self-diffusion coefficients of
charged suspensions are less affected by hydrodynamic interactions than
the corresponding coefficients of hard spheres.
As a major result we have found 
substantially different volume fraction dependencies of 
$H^t_s$ and $H^r_s$ for (deionized) charged and 
uncharged suspensions. The observed differences
are well explained in terms of an effective
hard sphere model by observing the big differences in the
microstructure of suspensions of charged and uncharged particles.

We note finally 
that recent DDLS measurements of $H^r_s$ in deionized suspensions
of charged fluorinated polymer particles compare favourably with our results
in eq. (\ref{hrs.charged}) \cite{Bitzer:private}. 
On the other hand, to our knowledge,
no experimental data
of $H^t_s$ for deionized charge-stabilized suspensions are accessable
so far.
We further point out that the interesting qualitative differences
between charge-stabilized suspensions and hard spheres exist also with
respect to sedimentation \cite{ThiesWeessie:95}
and long-time self-diffusion \cite{Naegele:Baur:forthcoming}.


\begin{thebibliography}{10}

\bibitem{Cichocki:88}
{B.~Cichocki} and {B.~U. Felderhof},
\newblock {\em J. Chem. Phys.} {\bf 89}, 1049 (1988).

\bibitem{Jones:1:88}
{R.~B. Jones},
\newblock {\em Physica} {\bf A 150}, 339 (1988).

\bibitem{Jones:Pusey:91}
{R.~B. Jones} and {P.~N. Pusey},
\newblock {\em Annu. Rev. Phys. Chem.} {\bf 42}, 137 (1991).

\bibitem{Beenakker:Mazur:83}
{C.~W.~J. Beenakker} and {P.~Mazur},
\newblock {\em Physica} {\bf A 120}, 388 (1983).

\bibitem{Degiorgio:95}
{V.~Degiorgio}, {R.~Piazza}, and {R.~B. Jones},
\newblock {\em Phys. Rev. E} {\bf 52}, 2707 (1995).

\bibitem{Naegele:Habil:published}
{G.~N{\"a}gele},
\newblock {\em Phys. Rep.} {\bf 272}, 215 (1996).

\bibitem{Watzlawek:96:1}
{M.~Watzlawek} and {G.~N{\"a}gele},
\newblock {\em Physica} {\bf A 235}, 56 (1997).

\bibitem{Watzlawek:forthcoming}
{M.~Watzlawek} and {G.~N{\"a}gele},
\newblock {\em Phys. Rev E} {\bf 56}, 1258 (1997).

\bibitem{Degiorgio:94}
{V.~Degiorgio}, {R.~Piazza}, and {T.~Bellini},
\newblock {\em Adv. Coll. Int. Sci.} {\bf 48}, 61 (1994).

\bibitem{Jones:Schmitz:88}
{R.~B. Jones} and {R.~Schmitz},
\newblock {\em Physica} {\bf A 149}, 373 (1988).

\bibitem{Bitzer:private}
{F.~Bitzer}, {T.~Palberg}, and {P.~Leiderer},
\newblock University of Konstanz,
\newblock private communication.

\bibitem{ThiesWeessie:95}
{D.~M.~E. Thies-Weessie}, {A.~P. Philipse}, {G.~N{\"a}gele}, {B.~Mandl}, and
  {R.~Klein},
\newblock {\em J. Coll. Int. Sci.} {\bf 176}, 43 (1995).

\bibitem{Naegele:Baur:forthcoming}
{G.~N{\"a}gele} and {P.~Baur},
\newblock {\em Europhys. Lett}, in press.

\end{thebibliography}
\end{document}